# Weblog patterns and human dynamics with decreasing interest

Jin-Li Guo

*Business School, University of Shanghai for Science and Technology, Shanghai 200093, China*

*Corresponding author. E-mail address: phd5816@163.com (J. –L. Guo)*

**Abstract** Weblog is the fourth way of network exchange after Email, BBS and MSN. Most bloggers begin to write blogs with great interest, and then their interests gradually achieve a balance with the passage of time. In order to describe the phenomenon that people's interest in something gradually decreases until it reaches a balance, we first propose the model that describes the attenuation of interest and reflects the fact that people's interest becomes more stable after a long time. We give a rigorous analysis on this model by non-homogeneous Poisson processes. Our analysis indicates that the interval distribution of arrival-time is a mixed distribution with exponential and power-law feature, that is, it is a power law with an exponential cutoff. Second, we collect blogs in ScienceNet.cn and carry on empirical studies on the interarrival time distribution. The empirical results agree well with the analytical result, obeying a special power law with the exponential cutoff, that is, a special kind of Gamma distribution. These empirical results verify the model, providing an evidence for a new class of phenomena in human dynamics. In human dynamics there are other distributions, besides power-law distributions. These findings demonstrate the variety of human behavior dynamics.

1. Introduction

Humans participate on a daily basis in a large number of distinct activities, from electronic communication to financial transactions. Human activities are too complex to be described well by one process or one model. In early 1920s, the counting process and the queueing theory were tried to apply to modeling human dynamics. The most widely used and, which is also the simplest, is a Poisson process. When Erlang studied a number of phone exchange calls, he discussed a Poisson distribution further and generalized it to an Erlang distribution. After that, the Poisson process is widely used in physics, communication, transportation, management science, etc.[1-2].

In 2005, Barabási analyzed the correspondence patterns of Einstein et al., data on the exchange of e-mails, and found that the timing of many human activities follows non-Poisson statistics.[3] Very long periods of inactivity that separate bursts of intensive activity follows a power-law distribution[3,4]. These findings challenge the application of Poisson processes in modeling of human activity. Poisson processes are characterized by a homogeneous activity pattern. More precisely, the time interval between two consecutive executions of a task follows an exponential distribution. However, the inter-event times of human activities are not always independent identically distributed. The empirical results of Barabási et al. indicate that human activity patterns are heterogeneous. It is characterized by a heavy tail in the distribution of the time interval between two consecutive executions of a given task. Malmgren et al. proposed a model of individual e-mail communication that is sufficiently rich to capture meaningful variability across individuals, while remaining simple enough to be interpretable. They found that variability of individual behavior over time is significantly less than variability across the population, suggesting that individuals can be classified into persistent "types"[7]. The mechanisms responsible for these marked features are diverse, such as executions of multiple tasks which are based on some perceived priority parameter [3-6] and interests on web browsing which are differ from person to person. All these can generate a power-law distribution of inter-event times [9, 11-15]. Yet, many empirical results indicate that inter-event times of human activities follow the power-law distribution with exponents 1 and 1.5, so Vázquez et al. divided human dynamics into two universality classes in Ref.[9].

In fact, human activities are quite complicated. A person may have greater interest in something as time goes on, for example, the frequency of drug taking may increase with the drug addiction. A person may also have less interest in something with the passage of time. People's interests in the same thing are diverse; some people have great interest in web browsing, while others have less interest in it. Thus, it's necessary to study features of the decreasing interest. Does the time interval distribution of events with the decreasing interest follow a power law? Can the interval distribution be characterized by a mixed distribution with exponential and power-law feature? If it follows a power law, does it belong to the two universality classes proposed by Vázquez et al.? Our goal in this paper is to address these questions.

The development of Internet provides a convenient platform for people to communicate with each other. It also provides a data source for the study of human behavior dynamics. The reason why human dynamic research becomes prominent in recent years is that large volume of data about human behaviors are now available. Blog is the abbreviation of Weblog (network diary). Weblog is the fourth network exchange way after Email, BBS and MSN. Most bloggers begin to write blogs with great interest, and their interests gradually achieve a balance over time. In Sec. 2, we propose a model of human dynamics with decreasing interest. This model describes attenuation in interest for a certain amount of time, by which people's interests get more stable with the passage of time. We give a rigorous analysis on this model by non-homogeneous Poisson processes. Our analysis indicates that the interval distribution of arrival-time is characterized by a mixed distribution with exponential and power-law property, that is, it is a power law with an exponential cutoff. In Sec. 3, we collect data on blogs in ScienceNet.cn and carry on an empirical study on interval distribution of the arrival-time of blogs. The empirical results agree well with the analytical result, obeying a special kind of Gamma distribution, that is, a special power law with the exponential cutoff. These empirical results verify the model, providing an evidence for a new class of phenomena in human dynamics. Conclusions are given in Sec. 4.

**2. Model and analysis**

Barabási model describes a kind of task selection pattern. This model assumes that the individual has a task list. According to the priority task, people determine the priority of processing tasks. The study of Barabási and Vázquez shows that the simple model can cause a fat tail distribution of task waiting time. However, not every sort of thing can be regarded as such tasks. We have an experience, when people are first in contact with new things their interests are often very great, however, as time goes on people's interest in these things gradually decays and achieves a balance. In order to describe the phenomenon of human behavior dynamic, we present a model as follows. (a) The number of events that occur in different time intervals is independent; (b) The rate of occurring event decreases linearly with time $t$, i.e., the rate at time $t$ is given by

$$\lambda(t) = \beta + \alpha/(bt+1), \tag{1}$$

where $b > 0, \alpha > 0, \beta \geq 0$; (c) The occurring probability of an event at time $t$ is $\lambda(t)dt$, and an event hardly occurs more than twice in a short time interval $dt$.

The rate of events $\lambda(t) = \beta + \alpha/(bt+1)$ shows that personal interest in an event diminishes over time. For a sufficiently large $t$, $\lambda(t) = \beta + \alpha/(bt+1) \to \beta$. It shows that personal interest in something is stabilizing.

The stochastic process $\{N(t), t \geq 0\}$ is said to be a counting process if $N(t)$ represents the total number of "events" that occur by time $t$. From this definition we see that for a counting process must satisfy: (ⅰ) $N(t) \geq 0$; (ⅱ) $N(t)$ is integer valued; (ⅲ) If $s < t$, then $N(s) \leq N(t)$; (ⅳ) For $s < t$, $N(t) - N(s)$ equals the number of "events" that occur the interval $(s, t]$.

Consider the counting process $N(t)$, and let us denote the time of the first event by $T_1$. Further, for $n > 1$, let $T_n$ denote the elapsed time between the (*n-1*)th and the *n*th event. The sequence $\{T_n, n = 1, 2, \cdots\}$ is called the sequence of interarrival times. Another quantity of interest is $S_n$, the arrival time of the *n*th event. It is easily seen that

$$S_0 = 0, \quad S_n = \sum_{i=1}^{n} T_i, \quad n = 1, 2, \cdots \tag{2}$$

The intensity function of the counting process $\{N(t), t \geq 0\}$ (if the defined limit exists) is given by

$$\lambda(t) = \lim_{\Delta t \to 0} \frac{P\{N(t + \Delta t) - N(t) > 0\}}{\Delta t} \tag{3}$$

It is a quantity that can well describe the character of a counting process. $\lambda(t)dt$ is the occurring probability of an event between $t$ and $t + dt$.

It is said to have independent increments if the number of events that occur in disjoint intervals are independent. It is said to have stationary increments if the distribution of the number of events that occur in any interval of time depend only on the length of the interval. The homogeneous Poisson process, as an important counting process, has stationary and independent increments, which is one of its fundamental features. If we eliminate the condition of stationary increments, we obtain a non-homogeneous Poisson process. According to Ref.[10], our model is a non-homogeneous Poisson process with intensity function $\lambda(t)$. From Ref.[2], two lemmas are introduced as follows.

**Lemma 1**. Let $\{N(t), t \geq 0\}$ be a non-homogeneous Poisson process with intensity function $\lambda(t)$. Then, for any given positive integer $n$ and $r$, the joint probability density function of $S_{r+1}, S_{r+2}, \cdots, S_{r+n}$ is given by

$$f_{S_{r+1}, S_{r+2}, \cdots, S_{r+n}}(t_{r+1}, t_{r+2}, \cdots, t_{r+n}) = \begin{cases} (\prod_{i=1}^{n} \lambda(t_{r+i})) \exp\{-\int_0^{t_{r+n}} \lambda(x)dx\} \frac{\left(\int_0^{t_{r+1}} \lambda(x)dx\right)^r}{r!} \\ \qquad \text{if } 0 < t_{r+1} \leq \cdots \leq t_{r+n} \\ 0 \qquad \qquad \qquad \qquad \qquad \text{otherwise} \end{cases} \tag{4}$$

**Lemma 2**. Let $\{N(t), t \geq 0\}$ be a non-homogeneous Poisson process with intensity function $\lambda(t)$. Then, for any given positive integer $n$, the conditional distribution function of the time interval $T_{n+1}$ between the *(n+1)*th and the *n*th events is given by

$$P\{T_{n+1} \leq t \mid S_n = x_n, \cdots, S_1 = x_1\} = 1 - \exp\{-\int_{x_n}^{x_n + t} \lambda(u)du\} \tag{5}$$

According to lemma 1 and lemma 2, we have

$$P\{T_{n+1} > t\} = \int_0^\infty dx_n \int_0^{x_n} dx_{n-1} \int_0^{x_{n-1}} dx_{n-2} \cdots \int_0^{x_3} dx_2 \int_0^{x_2} \exp\{-\int_0^{x_n+t} \lambda(u)du\} \prod_{i=1}^n \lambda(x_i) dx_1$$

$$= \int_0^\infty \exp\{-\int_0^{x_n+t} \lambda(u)du\} \lambda(x_n)dx_n \int_0^{x_n} \lambda(x_{n-1})dx_{n-1} \int_0^{x_{n-1}} \lambda(x_{n-2})dx_{n-2} \cdots \int_0^{x_3} \lambda(x_2)dx_2 \int_0^{x_2} \lambda(x_1)dx_1$$

$$= \int_0^\infty \exp\{-\int_0^{x_n+t}(\beta + \frac{\alpha}{bu+1})du\} \lambda(x_n)dx_n \int_0^{x_n} \lambda(x_{n-1})dx_{n-1} \cdots \int_0^{x_3} \lambda(x_2)dx_2 \int_0^{x_2} \lambda(x_1)dx_1$$

$$= \int_0^\infty \exp\{-(\beta x_n + \beta t + \frac{\alpha}{b}\ln(bx_n + bt + 1))\} \lambda(x_n)dx_n \int_0^{x_n} \lambda(x_{n-1})dx_{n-1} \cdots \int_0^{x_3} \lambda(x_2)dx_2 \int_0^{x_2} \lambda(x_1)dx_1$$

$$= e^{-\beta t} \int_0^\infty (bx_n + bt + 1)^{-\frac{\alpha}{b}} e^{-\beta x_n} \lambda(x_n)dx_n \int_0^{x_n} \lambda(x_{n-1})dx_{n-1} \cdots \int_0^{x_3} \lambda(x_2)dx_2 \int_0^{x_2} \lambda(x_1)dx_1$$

Hence, the distribution function of the interarrival time $T_{n+1}$ is given by

$$F(t) = 1 - P\{T_{n+1} > t\}$$

Here we aim to obtain a better understanding of the arrival-time interval distribution by doing asymptotic analysis to determine the limiting form as $t \to \infty$. Let

$$c = \int_0^\infty e^{-\beta x_n} \lambda(x_n)dx_n \int_0^{x_n} \lambda(x_{n-1})dx_{n-1} \cdots \int_0^{x_3} \lambda(x_2)dx_2 \int_0^{x_2} \lambda(x_1)dx_1.$$

Hence, we have the cumulative distribution of the interarrival time as follows

$$1 - F(t) \sim cb^{-\frac{\alpha}{b}}(t + \frac{1}{b})^{-\frac{\alpha}{b}} e^{-\beta t} \qquad \text{as} \qquad t \to \infty \qquad (6)$$

for positive constants $c$, $b$ and $\alpha$, where $f(t) \sim g(t)$ as $t \to \infty$ means that $f(t)/g(t) \to 1$ as $t \to \infty$; see Abate and Whitt [16]. The equation (6) shows that arrival-time interval distribution is a mixed distribution with exponential and power-law feature, that is, it is a power law with an exponential cutoff.

If $\alpha < b$, for sufficiently large $t$, the equation (6) also indicates that arrival-time interval distribution approximates to a Gamma distribution.

## 3. Empirical results

With the prevalence of the network technology and the Internet, many people become interested in writing Blogs. At the beginning, Blogers have a keep interest. But people's interest tends to be stable with the passage of time. In order to verify our model, we make statistic to the blog of four blogers in ScienceNet.cn. For convenience, A, B, C and D respectively represent bloger A, bloger B, bloger C and bloger D. The statistical data is in Table 1. In order to reduce the noise in the tail, we use the cumulative distribution. Figure 1 presents the blog arrival-time interval cumulative distribution in double-logarithmic coordinates. This figure shows that the empirical results agree with the analytical result well, obeying the approximately a Gamma distribution. In human dynamics there are other distributions, not just power-law distributions. These findings indicate the variety of human behavior dynamics.

**Tab. 1.** Data of blogs

| Bloger \ The number of blogs, time horizon | The number of blogs | Time horizon |
|---|---|---|
| A | 588 | 2007-3-15 to 2010-2-20 |
| B | 191 | 2007-7-11 to 2010-2-6 |
| C | 536 | 2008-3-7 to 2010-2-21 |
| D | 772 | 2007-1-20 to 2010-2-21 |

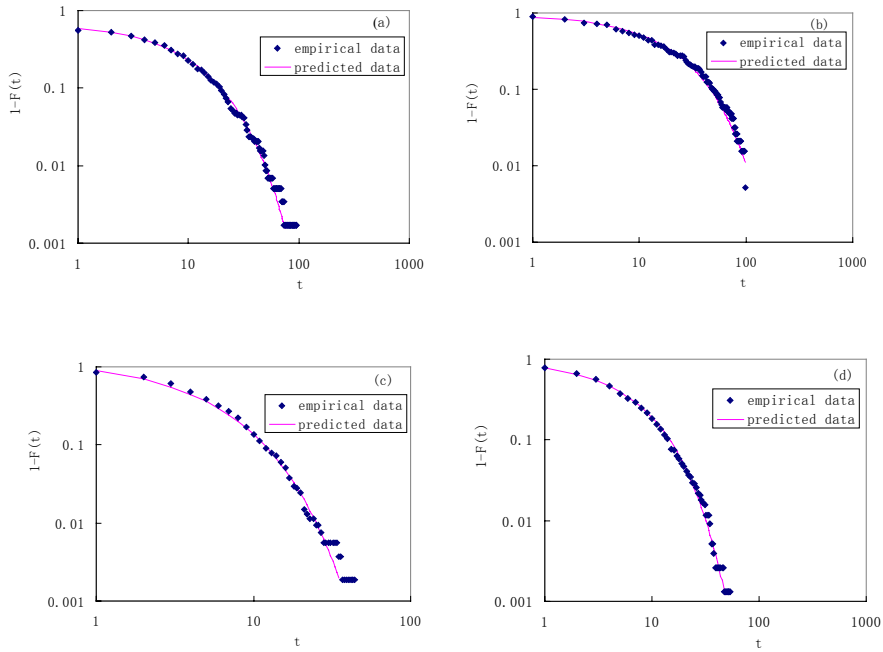

**Fig.1.** Double-logarithmic plot of the blog arrival-time interval cumulative distribution for four blogers. The horizontal axis for each panel is time and the vertical axis is the cumulative probability distribution of the blog arrival-time interval. ◆ empirical data; — predicted data by the model. (a) Empirical result of bloger A, theoretical cumulative probability distribution is given by $1-F(t)=1.85(t+7.7)^{-0.5}e^{-0.065t}$; (b) empirical result of bloger B, theoretical cumulative probability distribution is given by $1-F(t)=1.19(t+4)^{-0.16}e^{-0.04t}$; (c) empirical result of bloger C, theoretical cumulative probability distribution is given by $1-F(t)=1.95(t+3)^{-0.45}e^{-0.15t}$; (d) empirical result of bloger D, theoretical cumulative probability distribution is given by $1-F(t)=1.08(t+1.6)^{-0.2}e^{-0.125t}$. The empirical results agree with the analytical result well.

## 4. Conclusion

This paper investigates a phenomenon with decreasing interest. The empirical results show that blog arrival-time interval distribution approximates to the Gamma distribution and it is a mixed distribution with exponential and power-law feature. According to features of blogs, we establish a dynamic model of human-interest attenuation. This model has the following properties.

If $\alpha = 0, \beta > 0$, the time interval $T$ is an exponential random variable with parameter $\beta$, that is, the total number of "events" is a Poisson process having rate $\beta$.

If $\alpha > 0, \beta = 0$, as $b > 0$, the time interval $T$ follows a fat tailed distribution; as $b = 0$, the time interval $T$ is an exponential random variable with parameter $\alpha$, that is, the total number of "events" is a Poisson process having rate $\alpha$.

If $\alpha > 0, \beta > 0$, as $b > 0$, the arrival-time interval distribution approximates to a power law with an exponential cutoff; as $b = 0$, the time interval $T$ is an exponential random variable with parameter $\alpha + \beta$, that is, the total number of "events" is a Poisson process having rate $\alpha + \beta$.

From the above discussion, we can see that our model can describe a number of human dynamics with gradually decreasing interest. These results reflect the variety and complexity of human behavior with decreasing interest.

**Acknowledgments.** —This work was supported by the National Natural Science Foundation of China under Grant No.70871082 and the Foundation of Shanghai Leading Academic Discipline Project (Grant No. S30504). We would like to thank Dr. Xiao-Qing Jia for helpful in English.